\documentclass[letterpaper, 10 pt, conference]{ieeeconf}  

\IEEEoverridecommandlockouts                              

\usepackage{amsmath} 
\usepackage{amssymb}  
\usepackage{mathrsfs} 
\usepackage{graphicx,cite}
\usepackage{enumerate}
\usepackage{comment}
\usepackage{algorithm}
\usepackage{algpseudocode}
\usepackage{tcolorbox}
\usepackage{tabularx}
\usepackage{array}
\usepackage{dsfont}
\usepackage{setspace}

\newtheorem{rem}{Remark}
\newtheorem{prop}{Proposition}
\newtheorem{lem}{Lemma}

\newtheorem{defn}{Definition}

\newcommand{\Mcal}[1]{\mathcal{#1}}

\newcommand{\mnl}[1]{{\color{magenta} MNL: #1}}

\providecommand{\keywords}[1]{\textbf{\textit{Index terms---}} #1}

\newcommand{\remend}{\relax\ifmmode\else\unskip\hfill\fi\hbox{$\bullet$}}

\title{\LARGE \bf
Online Model-free Safety Verification for Markov Decision Processes Without Safety Violation
}

\author{Abhijit Mazumdar, Rafal Wisniewski and Manuela  L. Bujorianu 
\thanks{The work of the first and the second authors has been supported by the Poul Due Jensens Foundation under project SWIft.The work of the third author has been funded by the EPSRC project EP/R006865/1: Interface reasoning for interacting systems (IRIS).}
\thanks{A. Mazumdar and R. Wisniewski are with the Section of Automation $\&$ Control, Aalborg University, 9220 Aalborg East, Denmark (e-mail: \{abma, raf\}@es.aau.dk).}
\thanks{Manuela L. Bujorianu is with the Department of Computer Science, University College London, UK (e-mail: l.bujorianu@ucl.ac.uk).}}

\graphicspath{{figs/}}

\begin{document}

\maketitle
\thispagestyle{empty}
\pagestyle{empty}

\begin{abstract}

   In this paper, we consider the problem of safety assessment for Markov decision processes \textit{without} explicit knowledge of the model. We aim to learn probabilistic safety specifications associated with a given policy without compromising the safety of the process. To accomplish our goal, we characterize a subset of the state-space called \textit{proxy set}, which contains the states that are \textit{near} in a probabilistic sense to the \textit{forbidden set} consisting of all unsafe states. We compute the safety function using the single-step temporal difference method. To this end, we relate the safety function computation to that of the value function estimation using temporal difference learning. Since the given control policy could be unsafe, we use a safe \textit{baseline sub-policy} to generate data for learning. We then use an \textit{off-policy} temporal difference learning method with \textit{importance sampling} to learn the safety function corresponding to the given policy. Finally, we demonstrate our results using a numerical example.   
  \end{abstract}
  \vspace{0cm}
\section{Introduction}
In safety-critical systems, assessing safety associated with a control policy is crucial during the deployment of the control policy. Safety verification for dynamical system is usually studied in two settings: worst-case \cite{prajna2004safety,chutinan2003computational,sloth2012compositional} or stochastic \cite{bujorianu2003reachability,bujorianu2004extended,prajna2007framework,wisniewski2021safety,wisniewski2023probabilistic}. In the worse-case set up, safety corresponds to the property of never visiting the unsafe region. While the systems may be subjected to uncertain disturbance input, a hard upper bound on the disturbance input is assumed to be known. In the stochastic setup, safety is defined as the probability of reaching the unsafe region with a small probability below a prescribed margin. 
\par If the operational environment is changing or no prior information regarding the system model, or the environment is known, then safety needs to be verified during operation \cite{althoff2015online,gruber2018anytime,taye2022reachability}. This set-up is called \textit{online} safety verification. 
\par The works described above are model-based, i.e., an appropriate system model is required. Data-driven safety verification methods are getting attention, of late, as they eliminate the requirement of a model of the systems \cite{lavaei2021formal,salamati2021data,noroozi2021data,salamati2022data,salamati2022safety,mazumdar2023online}. Among these works, \cite{mazumdar2023online} considers a probabilistic safety notion, whereas \cite{lavaei2021formal,salamati2021data,noroozi2021data,salamati2022data,salamati2022safety} consider the worse-case safety definition. 
For systems with discrete-time and continuous states, \cite{salamati2021data,lavaei2021formal} proposed a data-driven method based on barrier certificate to verify safety formally. In \cite{noroozi2021data}, a data-driven approach with formal guarantees is presented for networks
of discrete-time sub-systems. To this end, a sub-barrier function for each sub-system is computed, then the overall barrier function is derived from the individual sub-barrier function. Further, \cite{salamati2021data, salamati2022data,salamati2022safety} converts the problem of finding barrier certificate as a robust convex problem. 
\\
\par \textit{Main Contributions:} In the existing works on data-driven safety verification, it is assumed that an existing data set is available. This is called the \textit{offline} set-up. However, many times, safety needs to be verified during the operating phase of a system in an \textit{online} fashion \cite{althoff2015online,gruber2018anytime,taye2022reachability}. If no prior data is available, the data-driven \text{online} set-up becomes more challenging as safety can be jeopardized during the learning. 
 \par To the best of our knowledge, the existing data-driven methods, except for \cite{mazumdar2023online}, are offline. In this paper, we develop an online safety verification method for stochastic systems without jeopardizing the system's safety. We consider a Markov decision process framework to represent the stochastic dynamics. Unlike \cite{mazumdar2023online}, in this work, we do not need to know even a partial model of the system. This relaxation makes the problem much harder compared to \cite{mazumdar2023online}. We use a single-step temporal difference method (TD($0$)) to learn the safety function corresponding to a given \textit{target control policy} $\pi$. If the TD($0$) method is used naively, then the target policy $\pi$, which needs to be assessed, must be used. However, since the policy $\pi$ is arbitrary and could be unsafe, employing it during the learning phase can lead to violation of the safety constraints. To circumvent this issue, we use an off-policy TD($0$) method with \textit{importance sampling} \cite{graves2022importance,precup2000eligibility}. We assume that at least one \textit{safe baseline sub-policy} for each state of a sub-set, called \textit{proxy set}, of the state-space is known. This assumption is an essential requirement in safe reinforcement learning. 
 The safe baseline sub-policy is needed to use the off-policy TD ($0$) method to learn the safety function without violating the safety constraints. 
\\
 \par The organization of the paper is as follows. In Section \ref{notation}, we set up the relevant notations. We present the system description and the problem formulation in Section \ref{problem_formulation}. The main results are described in Section \ref{safety_ver}. In Subsection \ref{learn_safety}, we presented the algorithm following a thorough discussion. With a numerical example, we demonstrate our results in Section \ref{sec_example}. Finally, in Section \ref{sec_conclusion}, we conclude the paper and highlight a future extension to this work.
 
\section{Notations}\label{notation}
We use $\mathcal{X}$ to represent a finite set of states, while $\mathcal{A}$ represents a finite set of actions. We construct the sample space $\Omega$ of all sequences of the form $\omega = (x_1,a_1,x_2,a_2, \hdots) \in (\mathcal{X} \times \mathcal{A})^{\infty}$ with $x_i \in \mathcal{X}$ and $a_i \in \mathcal{A}$. The sample space $\Omega$ is equipped with the $\sigma$-algebra $\mathcal{F}$ generated by coordinate mappings: $X_t(\omega) = x_t$ and $A_t(\omega) = a_t$.  With a slight abuse of notation, we shall use $X_t$ and $A_t$ to denote random variables, whereas $x_t$ and $a_t$ are deterministic values, their realizations, at time-step $t$. We suppose that $\mu$ is the distribution of the initial states $X_0$. 
In this work, we consider stationary policies, i.e., maps $\pi: \mathcal{X} \to \Delta(\mathcal{A})$, with $\Delta(\mathcal{A}) = \{(p_1, \hdots, p_{|\mathcal{A}|}) \in [0,1]^{\mathcal{A}} \mid p_1 + \hdots + p_{|\mathcal{A}|} = 1\}.$ A sub-policy $\pi'$ for a subset of $W\subseteq \mathcal{X}$ is defined as $\pi': W \to \Delta(\mathcal{A})$. For a fixed initial distribution $\mu$ and a policy $\pi$,  
we define recursively the probability $\mathds{P}^{\mu}_\pi$ on $\mathcal{F}$ by
\begin{align*}
   \mathds{P}^{\mu}_\pi [X_1 = x] &= \mu(x) \\
   \mathds{P}^{\mu}_\pi [A_t = a \mid X_t = x] &= \pi(a|x) \\
\mathds{P}^{\mu}_\pi [X_{t+1} = y \mid X_t = x, A_t = a] &= p({x,a,y})
\end{align*}
\par Specifically, the process $(X_t)$ with stationary policy is characterized by the transition probability $p_{x,y} = \sum_{a \in \mathcal{A}} p_{x,a,y} \pi(a|x)$. Hence, the process $(X_t)$ is time-homogeneous. 
We write $\mathds{P}^{y}_\pi := \mathds{P}^{\delta_y}_\pi$ for the delta distribution concentrated at $y$. The expectation with respect to $\mathds{P}^{y}_\pi$ is denoted $\mathds{E}^{y}_\pi$.
\section{System Description and Problem Formulation} \label{problem_formulation}
 We consider an MDP with a set of states $\mathcal{X}$ and a set of actions $ \mathcal{A}$. Suppose the set of states is partitioned into a target set $E \subset \mathcal{X}$, a set of forbidden states $U$, and $H := \mathcal{X} \setminus (E\cup U)$ be the set of living (taboo) states. 
\par 
\par This work deals with probabilistic safety. For any {target control policy}, or simply target policy $\pi$, in order to assess safety, a safety function $S_{\pi}(x)$ is defined as follows \cite{wisniewski2023probabilistic}.
\begin{defn}(Safety Function)
  For each state $x\in H$, the safety function is the  probability that the realizations hit the forbidden set $U$ before reaching the target set $E$, i.e., for a fixed policy $\pi$, 
  \begin{align*}
    S_{\pi}(x) := \mathds{P}^{x}_{\pi}[\tau_U < \tau_{E}],
\end{align*}
where  $\tau_A$ is the first hitting time of a set $A$.\remend
\end{defn}
\par We consider a probabilistic safety notion called $p$-safe \cite{wisniewski2021safety,wisniewski2023probabilistic}. Following definitions are central to this work and are inspired from \cite{wisniewski2021safety,wisniewski2023probabilistic}. 
\begin{defn} ($p$-Safe State, $p$-Safe MDP and $p$-Safe Policy) 
 For a given policy $\pi$, a state $x\in H$ is called $p$-safe if the safety function does not exceed $p$, i.e., $S_{\pi}(x)\leq p$. Similarly, an MDP is called $p$-safe with a policy $\pi$ if: $\underset{x\in H}{\text{max}} \text{ } S_{\pi}(x)\leq p$. In this case, $\pi$ is called a $p$-safe policy.  \remend
\end{defn}
\par It should be noted that we use \text{safety} and $p$-\textit{safety} synonymously throughout the paper.
\par We now formally state the problem that we address in this work as follows. 
\begin{tcolorbox}[colback=white!3,colframe=black!50!black,title= ]
 \text{\textit{Problem $P$:}} Estimate the safety function for the given \textit{target policy} $\pi$ without rendering the MDP unsafe, i.e., ensuring that $S_{\pi}(x)\leq p$ for each state $x\in H$.  
\end{tcolorbox}

\par Since the target policy, $\pi$, could be arbitrary; if we apply it, we might jeopardize the safety of the MDP. Hence, we must use an indirect way to learn about the safety function corresponding to $\pi$. 
 In order to solve Problem $P$, we now introduce a \textit{proxy set} as follows. 
\begin{defn}(Proxy Set)\label{Def.ProxySet}
We call the subset $U'\in H$ as a \textit{proxy set} of an MDP, if it has the following properties:
\begin{enumerate}[N.1]
\item  $\tau_{U'}<\tau_U$, almost surely. 
\item For all $x\in U'$, there exists $a\in A$ and $y\notin U$ such that $p(x,a,y)>0$. \remend
\end{enumerate}
\end{defn}
 \par The proxy set $U'$ can be considered a neighborhood of the forbidden set $U$ as the probability of hitting $U'$ before hitting the forbidden set $U$ is $1$.
 \begin{figure}[!htb] 
  \includegraphics[scale=0.5]{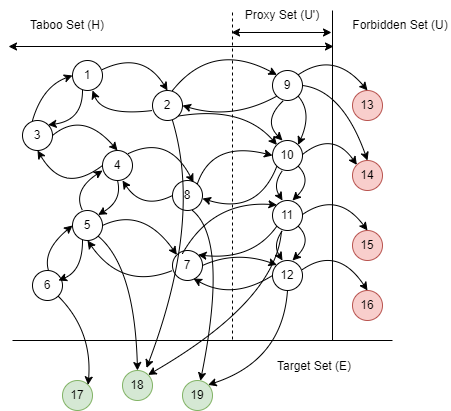}
\caption{Pictorial description of the taboo set, proxy set, forbidden set, and target set.}
\label{fig:r_avg}
\end{figure}
 \par We now introduce the concept of \textit{safe action} and \textit{safe baseline sub-policy} for the proxy states $x'\in U'$. These will enable us to learn the safety function of a given policy without violating the safety constraint.
 \begin{defn}(Safe Action)
     For each proxy state $x'\in U'$, we call an action a safe action, denoted by $a^{safe}(x')$, if 
\[
p(x',a^{safe}(x'),y)=0 \text{, } \forall y\in U.
\]     
\remend
 \end{defn}
\begin{defn}(Safe Baseline Sub-Policy)
We call a sub-policy defined for the proxy set $U'$ a \textit{safe baseline sub-policy}, denoted by $\pi^{S}$, if
\[
S_{\pi^S}(x')\leq p \text{, }\forall x'\in U'.
\]
\remend
\end{defn}
\par Throughout the paper, we have the following assumptions:
\begin{enumerate}[i)]
    \item The proxy set $U'$ is given.
    \item A safe base sub-policy $\pi^{S}$,  for each proxy state $x'\in U'$, is known.
\end{enumerate}
    \section{Online Safety Verification} \label{safety_ver}
    In this section, we present the algorithm for online safety verification. Before presenting the algorithm for estimating the safety function, we need to establish the following results.
\par As given in \cite{wisniewski2023probabilistic}, safety function can be expressed as follows. 
\begin{lem} (\cite{wisniewski2023probabilistic}) \label{lem_safe_kappa}
Suppose, $\tau = \tau_{U \cup E}$ is almost surely finite. The safety function for each state $x\in H$, with a policy $\pi$, is then given by
\begin{equation*}\label{SafetyCummulative}
    S_{\pi}(x) = \mathds{E}_{\pi}^{x} \sum_{t = 0}^{\tau-1} \kappa(X_t, A_t),
\end{equation*}
where $\kappa(x,a) = \sum_{y \in U} p({x,a,y})$. \remend
\end{lem}
\begin{rem}
     If we knew the transition probabilities $p({x,a,y})$ from the proxy set to the forbidden set, as in \cite{mazumdar2023online}, we could use the standard TD($0$) method to estimate the safety function. Since we do not have access to the transition probabilities, we need to express the safety function as follows.  \remend
\end{rem}   
\begin{prop}\label{safety_fun_RL}
 If $\tau = \tau_{U \cup E}< \infty$, almost surely, then the safety function for any state $x\in H$ can be expressed as follows:
\begin{align}\label{SafetyCummulative_new}
    S_{\pi}(x) = \mathds{E}_{\pi}^x \sum_{t = 0}^{\tau-1} c(X_t, A_t),
\end{align}
where,
\begin{equation}
    \begin{split} 
   & c(X_t,A_t) = \begin{cases}
			 1, & \text{if } X_{t+1}\in U\\
            0, & \text{otherwise.} 
		 \end{cases} \\
\end{split}
\label{cost}
\end{equation}
\end{prop}
\begin{proof}
    Observe the following:
    \begin{equation}
        \begin{split}
           & \mathds{E}_{\pi}^x [c(X_t,A_t)|X_t=\tilde{x},A_t=a] \\
            & = \mathds{E}_{\pi}^x \big[\mathds{E}_{\pi} [c(X_t,A_t)|X_t=\tilde{x},A_t=a]\big] \\
            & = \mathds{E}_{\pi}^x [ \sum_{y\in U} p(\tilde{x},a,y)]\\
            & = \mathds{E}_{\pi}^x [ \kappa(X_t,A_t)|X_t=\tilde{x},A_t=a]
        \end{split}
    \end{equation}
    Thus, the expressions of the safety function given in Lemma \ref{lem_safe_kappa} and Proposition \ref{safety_fun_RL} are equivalent. Hence, using Lemma \ref{lem_safe_kappa}, the safety function can be expressed as given in the Proposition.
\end{proof}

\par The following property relates the safety function of proxy states $x'\in U'$ with the safety function of states $x\in H\setminus U'$. This property will be used to ensure safety during the learning phase.
\begin{prop}\label{lem_safe_proxy} 
    For any state $x\in H$, the following is true:
    \begin{equation*}
       S_{\pi}(x) \leq \underset{x'\in U'}{\text{max}} \ S_{\pi}(x').
    \end{equation*}
\end{prop}
\begin{proof}
Following the definition of the safety function, for any $x\in H$, we get the following:
\begin{equation*}
    \begin{split}
        S_{\pi}(x) & = \mathds{P}_{\pi}^x [\tau_U < \tau_E] \\
        & = \sum_{x' \in U'} \mathds{P}_{\pi}^{x'} [\tau_U < \tau_E] \mathds{P}_{\pi}^{x} [X_{\tau'}= x'] \ (\text{let } \tau' = \tau_{U'\cup E}) \\
        & =  \sum_{x' \in U'} S_{\pi}(x') \mathds{P}_{\pi}^{x} [X_{\tau'}= x'] \\
     &  \leq \underset{x'\in U'}{\text{max}} \ S_{\pi}(x').
    \end{split}
\end{equation*}
The second equality is a direct consequence of the first property of the proxy states.  
    \end{proof}
\begin{rem}\label{rem1}
    As a consequence of Proposition \ref{lem_safe_proxy}, if we apply a safe baseline policy only for the proxy states $U'$ and use a target policy $\pi$ for other states $H\setminus U'$, the MDP will be safe throughout the learning phase. \remend
\end{rem}
\subsection{Safe learning of the safety function} \label{learn_safety}
\par We notice that the safety function in \eqref{safety_fun_RL} resembles the value function considered in reinforcement learning. The single-step temporal difference method, TD($0$), is one of the most widely used methods to compute the value function.  Hence, we also use the TD($0$) method to estimate the safety function. Suppose ${S}_t(x)$ is the estimated safety function for state $x$ in the $t${th} learning step, and after applying $A_t$ according to $\pi$ the process reaches state $y$. Then, according to the TD($0$) method, the update rule for the estimated safety function is as given below:
\begin{equation}
\begin{split}  
  &  {S}_{t+1}(x) \leftarrow {S}_{t}(x) + \alpha_t(x) [c_t+ {S}_{t}(y)-{S}_t(x)], \\
          & \text{where, } c_t = \begin{cases}
			 1, & \text{if } y\in U\\
            0, & \text{if } y\notin U
		 \end{cases}. 
\end{split}
\label{reward_eq}
\end{equation} 
In the above expression, $c_t+S_t(y)$ is called the \textit{TD target}. 
\par Now, suppose the learning rate $\alpha_t(x)$ is chosen such that following conditions are satisfied: 
\begin{equation}
    \begin{split}
     & (i) \sum_{t}^{\infty} \alpha_t(x) = \infty \\
     & (ii) \sum_{t}^{\infty} \alpha^2_t(x) < \infty.
    \end{split}
    \label{learn_rate}
\end{equation}
Then, using the results presented in \cite{tsitsiklis1994asynchronous}, it can be inferred that ${S}_{t}(x)$ converges to the true safety function $S(x)$ for each $x\in H$.
 \par Since the hitting time is finite (almost surely), we consider an episodic temporal difference TD($0$) algorithm. In an episodic learning framework, whenever the process hits the terminal states, learning is resumed from an arbitrary initial state. Since the given target policy $\pi$ could render the MDP unsafe, we use a safe baseline sub-policy $\pi^S$ for the proxy set to generate data. However, the goal is to learn the safety function $S_{\pi}(x)$ with the target policy $\pi$. Since $\pi^S$ is safe by definition, the safety is always maintained for the proxy set $U'$. Further, if we ensure that the safety function for the proxy set $U'$ is less than $p$, then from Proposition \ref{lem_safe_proxy}, it is made sure that the MDP is safe. The policy that is used to generate necessary data during learning is called the \textit{behavior policy}, denoted by $\pi^b$. The \textit{behavior policy} $\pi^b$ that we use is as follows:
\begin{equation}
    \pi^b = \begin{cases}
			 \pi, & \text{ for } x \in H \setminus U'\\
           \pi^{S}  & \text{ for } x \in U'.
		 \end{cases}
   \label{behav_pol_1}
\end{equation}
Since the behavior policy $\pi^b$ is chosen differently than the target policy $\pi$, if we use the standard TD($0$) naively, then we would only learn $S_{\pi^b}(x)$, and not $S_{\pi}(x)$. To resolve this issue, we use a variant of TD($0$) with \textit{per-decision importance sampling}, which is an \textit{off-policy} value function estimation method as given in \cite{graves2022importance}. In this method, for the proxy states $x'\in U'$, the update rule for the estimated safety function takes the following form:
\small
\begin{equation}
      {S}_{t+1}(x') \leftarrow {S}_{t}(x') + \alpha_t(x') [\frac{\pi(a|x')}{\pi^S(a|x')} \big(c_t+ {S}_{t}(y)-{S}_t(x')\big)],
\end{equation}
\normalsize
where, $c_t$ is as given in \eqref{reward_eq} and $\frac{\pi(a|x')}{\pi^S(a|x')}$ is called the \textit{importance-sampling ratio}. The learning rate is kept similar to \eqref{learn_rate}. From the results presented in \cite{precup2000eligibility}, it follows that $S_{t+1}(x')$ converges to the true safety function $S_{\pi}(x')$ with the target policy $\pi$, almost surely.
\par From Proposition \ref{lem_safe_proxy}, it is clear that if we ensure that the proxy states $x'\in U'$ are safe, then the states $x\in H\setminus U'$ will be safe irrespective of the sub-policy used for them. However, since we use the policy $\pi$ that needs to be assessed for any state $x\in H\setminus U'$, the update rule is the standard one as given in \eqref{reward_eq}.  
\par Further, to estimate the safety function with the behavior policy $\pi^b$, the update rule is given by: 
\begin{equation}
\begin{split}  
  &  {S}^b_{t+1}(x) \leftarrow {S}^b_{t}(x) + \alpha_t(x) [c_t+ {S}^b_{t}(y)-{S}^b_t(x)], \\
          & \text{where, } c_t = \begin{cases}
			 1, & \text{if } y\in U\\
            0, & \text{if } y\notin U
		 \end{cases}. 
\end{split} 
\label{reward_eq_1} 
\end{equation} \remend
\small
\begin{algorithm}
  \caption{: Safe TD($0$) algorithm for estimating the safety function:}
  \begin{algorithmic}[1]
    \State \textbf{Input:} The given target policy $\pi$ for which safety is needed to be evaluated, a safe baseline sub-policy $\pi^S$, a safe behavior policy $\pi^b$, learning rate $\alpha_t(x)$ for each $x\in H$, safety parameter $p$, proxy set $U'$; 
    \State \textbf{Initialize:} ${S}_1(x)$ for each $x\in H$ arbitrarily, ${S}_1(x)=0$ for each $x\in U\cup E$, $t=1$; 
    \For  {Episodes ($k=1,2,...,\mathcal{L}$)}
    \State Draw an initial state $x$ uniformly from $H$;
    \For  { Iterations ($i=1,2,...,\mathcal{T}$)} 
    \If{ $x\in U'$ }
     \State Apply a safe action $A_t=a^{safe}(x)$ according to the safe baseline sub-policy $\pi^S$;
    \Else
    \State Apply action $A_t$ according to the target policy $\pi$;
    \EndIf
    \State Observe the new state $y$, and $c_t$ according to \eqref{reward_eq}; 
    \If{$x\in U'$}
    \State Update the safety function as follows:
    \color{black}
    \small
    \begin{equation}
        \begin{split}
           {S}_{t+1}(x) \leftarrow {S}_{t}(x) + \alpha_t(x) [\frac{\pi(a|x)}{\pi^S(a|x)} \big(c_t+ {S}_{t}(y)-{S}_t(x)\big)];
        \end{split}
    \end{equation}
    \normalsize
    \Else \ ($x\in H\setminus U'$)
    \State Update the safety function as follows:
    \small
    \begin{equation}
        \begin{split}
           {S}_{t+1}(x) \leftarrow {S}_{t}(x) + \alpha_t(x) [c_t+ {S}_{t}(y)-{S}_t(x)];
        \end{split}
    \end{equation}
    \normalsize
    \EndIf
    \color{black}
    \State Set $x \leftarrow y$;
    \State Set $t \leftarrow i+1$;
     \If{$x$ is a terminal state, i.e., $x\in U \cup E$}
    \State Terminate the Episode. 
    \EndIf
     \EndFor
     \color{black}
     \color{black}
     \EndFor
  \end{algorithmic}
\end{algorithm}
\normalsize

 \section{Illustrating Example} \label{sec_example}
\begin{figure}[!htb] 
  \includegraphics[scale=0.5]{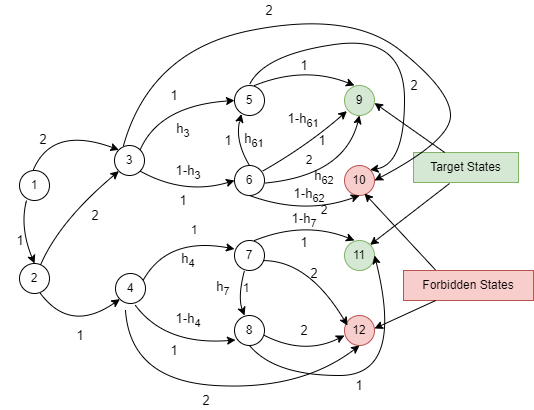}
\caption{Example MDP.}
\label{fig:example}
\end{figure}

 Consider the MDP as shown in Figure \ref{fig:example}. 
 There are $12$ states, of which two are forbidden, and two are target states. Specifically, the set of states is $\mathcal{X}=\{1,2,...,12\}$, target set is $E=\{9,11\}$, forbidden set is $U=\{10,12\}$, the taboo set is $H=\{1,2,...,8\}$, and the set of actions is $\mathcal{A}=\{1,2\}$.  
\par
\begin{figure}[!htb] 
  \includegraphics[scale=0.5]{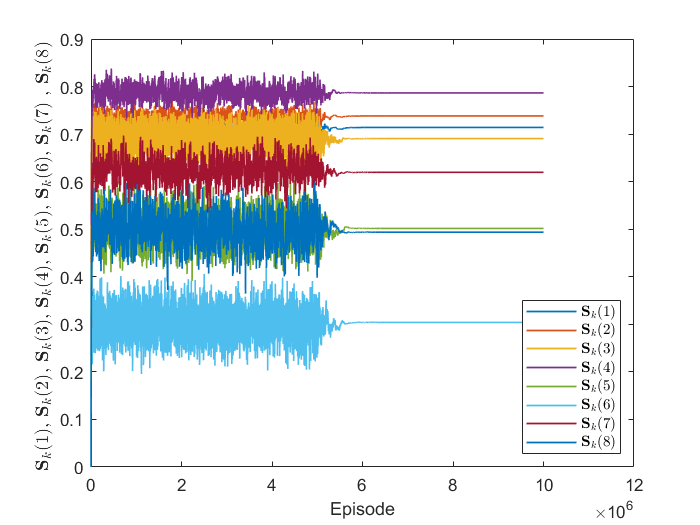}
\caption{Convergence of the safety function with the target policy $\pi$.}
\label{fig:safety}
\end{figure}

\begin{figure}[!htb] 
  \includegraphics[scale=0.5]{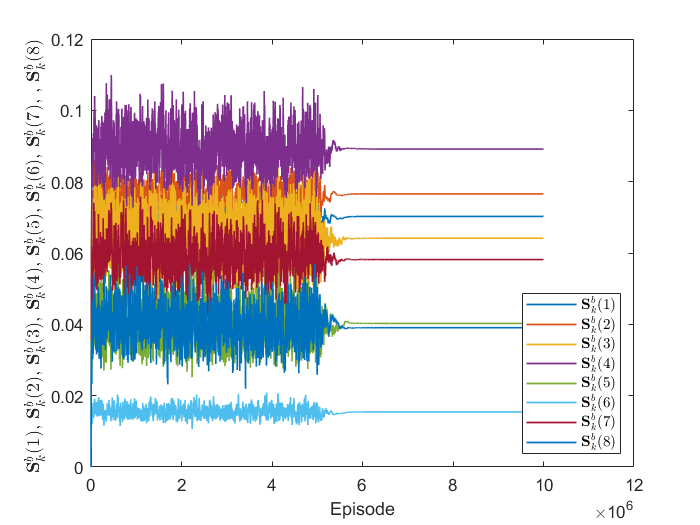}
\caption{Convergence of the safety function with the behavior policy $\pi^b$.}
\label{fig:safety_safe}
\end{figure}

 While we do not need any model parameters, i.e., the transition probabilities, we assume the proxy set is known. In the above example, the proxy set is $U'=\{3,4,5,6,7,8\}$. Suppose we are given to assess the safety function with the target policy $\pi$, which is a uniformly random policy for each state, i.e., $\pi(a|x)=0.5$ for each $x\in H$ and $a\in \mathcal{A}$. Assume that the MDP must be $p$-safe with $p=0.1$. We assume that a safe policy is known from each proxy state $x'\in \{3,4,5,6,7,8\}$. 
For each $x'\in U'$, a safe sub-policy is as follows:
\begin{equation*}
    \begin{split}
        \pi^S(a|x') = \begin{cases}
               {0.96}, & \text{if } a= 1 \\
               {0.04}, & \text{if } a=2,
                \end{cases}
    \end{split}
\end{equation*}
 Assume that $h_3=0.4$, $h_4=0.6$, $h_{61}=0.4$, $h_{62}=0.6$ and $h_7=0.5$. We show the convergence of the estimated safety function with the target policy and the behavior policy in Figure \ref{fig:safety} and \ref{fig:safety_safe}, respectively. From the figures, it can be seen that the safety function with the safe behavior policy is less than $p$ for all states, hence $p$-safe. Learning rate $\alpha_k(x)$ is chosen as follows:
 \begin{equation*}
     \begin{split}
         \alpha_k = \begin{cases}
             0.001, & \text{ for all episodes } k\leq \mathcal{L}/2 \\
             \frac{\alpha_{k-1}}{1 + (10^{-6} \cdot \text{log}(k+1)}), & \text{ for all episodes } k > \mathcal{L}/2.
         \end{cases}
     \end{split}
 \end{equation*}
 \par In Table I, we have shown the true value of the safety function for the target policy $\pi$ and the behavior policy. These are estimated using the result given in \cite{wisniewski2023probabilistic}. Further, the estimated safety function for the policies $\pi$ and $\pi^b$, at the end of the last episode $\mathcal{L}= 10^7$, are shown. It is observed that the final estimated safety functions $S_{\mathcal{L}}(x)$ and $S^b_{\mathcal{L}}(x)$ approach arbitrary close to the actual values.

\begin{table}[!htb]
\begin{center}
\begin{tabular}{ | m{2em} | m{1cm} | m{1.5cm} | m{1cm} | m{1.5cm} | } 
  \hline
  State ($x$) & True $S_{\pi}(x)$ using \cite{wisniewski2023probabilistic} &  ${S}_{\Mcal{L}}(x)$ using Algorithm 1 ($\mathcal{L}= 10^{7}$) & True $S_{\pi^b}(x)$ using \cite{wisniewski2023probabilistic}   & ${S}^b_{\Mcal{L}}(x)$ using Algorithm 1 ($\mathcal{L}= 10^{7}$) \\ 
  \hline
  1 & 0.7144  & 0.7140 & 0.0882 & 0.0703  \\ 
  \hline
  2 & 0.7387 & 0.7381 & 0.0888 & 0.0766  \\ 
  \hline
  3 & 0.69  & 0.6906 & 0.0734 & 0.0641 \\ 
  \hline
  4 & 0.7875  & 0.7866 & 0.0895 & 0.0891 \\
  \hline
  5 & 0.5  & 0.5017 & 0.04 & 0.0403 \\ 
  \hline
  6 & 0.3  & 0.3041 & 0.0314 & 0.0155 \\ 
  \hline
  7 & 0.625  & 0.6197 & 0.0592 & 0.0582 \\ 
  \hline
  8 & 0.5  & 0.4939 & 0.04 & 0.0391 \\ 
  \hline
\end{tabular}
\caption{\label{safety_bound-table} Comparison of the actual and estimated  safety function.}
\end{center}
\label{table_1}
\end{table}
\section{Conclusion and Future Work} \label{sec_conclusion}
We have presented a TD($0$) method for estimating the safety function without jeopardizing the safety of the MDP. Specifically, we have used an \textit{off-policy} TD ($0$) with \textit{per-decision importance sampling} to estimate the safety function. We have demonstrated that the estimated safety functions for all the states converge to the true value of the safety function. Further, we have shown that the safety functions with the \textit{behavior policy}, that is followed during the learning, also converges to their true values.
\par We are working on extending these results to MDP with a large number of states, and continuous dynamical systems. To this end, we will use function approximation-based reinforcement learning techniques.
\addtolength{\textheight}{-12cm}   





%
%
%
%
%
%
\bibliographystyle{IEEEtran}
\bibliography{ecc24}

\begin{thebibliography}{10}
\providecommand{\url}[1]{#1}
\csname url@samestyle\endcsname
\providecommand{\newblock}{\relax}
\providecommand{\bibinfo}[2]{#2}
\providecommand{\BIBentrySTDinterwordspacing}{\spaceskip=0pt\relax}
\providecommand{\BIBentryALTinterwordstretchfactor}{4}
\providecommand{\BIBentryALTinterwordspacing}{\spaceskip=\fontdimen2\font plus
\BIBentryALTinterwordstretchfactor\fontdimen3\font minus
  \fontdimen4\font\relax}
\providecommand{\BIBforeignlanguage}[2]{{%
\expandafter\ifx\csname l@#1\endcsname\relax
\typeout{** WARNING: IEEEtran.bst: No hyphenation pattern has been}%
\typeout{** loaded for the language `#1'. Using the pattern for}%
\typeout{** the default language instead.}%
\else
\language=\csname l@#1\endcsname
\fi
#2}}
\providecommand{\BIBdecl}{\relax}
\BIBdecl

\bibitem{prajna2004safety}
S.~Prajna and A.~Jadbabaie, ``Safety verification of hybrid systems using
  barrier certificates,'' in \emph{International Workshop on Hybrid Systems:
  Computation and Control}.\hskip 1em plus 0.5em minus 0.4em\relax Springer,
  2004, pp. 477--492.

\bibitem{chutinan2003computational}
A.~Chutinan and B.~H. Krogh, ``Computational techniques for hybrid system
  verification,'' \emph{IEEE transactions on automatic control}, vol.~48,
  no.~1, pp. 64--75, 2003.

\bibitem{sloth2012compositional}
C.~Sloth, G.~J. Pappas, and R.~Wisniewski, ``Compositional safety analysis
  using barrier certificates,'' in \emph{Proceedings of the 15th ACM
  international conference on Hybrid Systems: Computation and Control}, 2012,
  pp. 15--24.

\bibitem{bujorianu2003reachability}
M.~L. Bujorianu and J.~Lygeros, ``Reachability questions in piecewise
  deterministic markov processes,'' in \emph{Hybrid Systems: Computation and
  Control: 6th International Workshop, HSCC 2003 Prague, Czech Republic, April
  3--5, 2003 Proceedings 6}.\hskip 1em plus 0.5em minus 0.4em\relax Springer,
  2003, pp. 126--140.

\bibitem{bujorianu2004extended}
M.~L. Bujorianu, ``Extended stochastic hybrid systems and their reachability
  problem,'' in \emph{Hybrid Systems: Computation and Control: 7th
  International Workshop, HSCC 2004, Philadelphia, PA, USA, March 25-27, 2004.
  Proceedings 7}.\hskip 1em plus 0.5em minus 0.4em\relax Springer, 2004, pp.
  234--249.

\bibitem{prajna2007framework}
S.~Prajna, A.~Jadbabaie, and G.~J. Pappas, ``A framework for worst-case and
  stochastic safety verification using barrier certificates,'' \emph{IEEE
  Transactions on Automatic Control}, vol.~52, no.~8, pp. 1415--1428, 2007.

\bibitem{wisniewski2021safety}
R.~Wisniewski and L.-M. Bujorianu, ``Safety of stochastic systems: An analytic
  and computational approach,'' \emph{Automatica}, vol. 133, p. 109839, 2021.

\bibitem{wisniewski2023probabilistic}
R.~Wisniewski and M.~L. Bujorianu, ``Probabilistic safety guarantees for
  {M}arkov decision processes,'' \emph{IEEE Transactions on Automatic Control},
  2023.

\bibitem{althoff2015online}
D.~Althoff, M.~Althoff, and S.~Scherer, ``Online safety verification of
  trajectories for unmanned flight with offline computed robust invariant
  sets,'' in \emph{2015 IEEE/RSJ International Conference on Intelligent Robots
  and Systems (IROS)}.\hskip 1em plus 0.5em minus 0.4em\relax IEEE, 2015, pp.
  3470--3477.

\bibitem{gruber2018anytime}
F.~Gruber and M.~Althoff, ``Anytime safety verification of autonomous
  vehicles,'' in \emph{2018 21st International Conference on Intelligent
  Transportation Systems (ITSC)}.\hskip 1em plus 0.5em minus 0.4em\relax IEEE,
  2018, pp. 1708--1714.

\bibitem{taye2022reachability}
A.~G. Taye, J.~Bertram, C.~Fan, and P.~Wei, ``Reachability based online safety
  verification for high-density urban air mobility trajectory planning,'' in
  \emph{AIAA AVIATION 2022 Forum}, 2022, p. 3542.

\bibitem{lavaei2021formal}
A.~Lavaei, A.~Nejati, P.~Jagtap, and M.~Zamani, ``Formal safety verification of
  unknown continuous-time systems: a data-driven approach,'' in
  \emph{Proceedings of the 24th International Conference on Hybrid Systems:
  Computation and Control}, 2021, pp. 1--2.

\bibitem{salamati2021data}
A.~Salamati, A.~Lavaei, S.~Soudjani, and M.~Zamani, ``Data-driven verification
  and synthesis of stochastic systems through barrier certificates,''
  \emph{arXiv preprint arXiv:2111.10330}, 2021.

\bibitem{noroozi2021data}
N.~Noroozi, A.~Salamati, and M.~Zamani, ``Data-driven safety verification of
  discrete-time networks: A compositional approach,'' \emph{IEEE Control
  Systems Letters}, vol.~6, pp. 2210--2215, 2021.

\bibitem{salamati2022data}
A.~Salamati and M.~Zamani, ``Data-driven safety verification of stochastic
  systems via barrier certificates: A wait-and-judge approach,'' in
  \emph{Learning for Dynamics and Control Conference}.\hskip 1em plus 0.5em
  minus 0.4em\relax PMLR, 2022, pp. 441--452.

\bibitem{salamati2022safety}
------, ``Safety verification of stochastic systems: A repetitive scenario
  approach,'' \emph{IEEE Control Systems Letters}, vol.~7, pp. 448--453, 2022.

\bibitem{mazumdar2023online}
A.~Mazumdar, R.~Wisniewski, and M.~L. Bujorianu, ``Online learning of safety
  function for {M}arkov decision processes,'' in \emph{European Control
  Conference (ECC), 2023}.\hskip 1em plus 0.5em minus 0.4em\relax IEEE, 2023,
  pp. 1--6.

\bibitem{graves2022importance}
E.~Graves and S.~Ghiassian, ``Importance sampling placement in off-policy
  temporal-difference methods,'' \emph{arXiv preprint arXiv:2203.10172}, 2022.

\bibitem{precup2000eligibility}
D.~Precup, R.~S. Sutton, and S.~P. Singh, ``Eligibility traces for off-policy
  policy evaluation,'' in \emph{Proceedings of the Seventeenth International
  Conference on Machine Learning}, 2000, pp. 759--766.

\bibitem{tsitsiklis1994asynchronous}
J.~N. Tsitsiklis, ``Asynchronous stochastic approximation and {Q}-learning,''
  \emph{Machine learning}, vol.~16, pp. 185--202, 1994.

\end{thebibliography}
\end{document}